\documentclass[prb,twocolumn,superscriptaddress,showpacs]{revtex4}
\usepackage{graphicx}
\usepackage{color}

\usepackage[tbtags]{amsmath}

\usepackage[
pdftitle={Quantum tunneling of semifluxons},
pdfauthor={Dr. E. Goldobin},
pdfsubject={..},
urlcolor=blue,
bookmarks=true,
bookmarksopen=true,
bookmarksnumbered=true
]{hyperref}

\include{MyMc}
\newcommand{\da}{\delta\tilde{a}}

\begin{document}


\title{Quantum tunneling of semifluxons}

\author{E.~Goldobin}
\email{gold@uni-tuebingen.de}
\affiliation{
  Physikalisches Institut II,
  Universit\"at T\"ubingen,
  Auf der Morgenstelle 14,
  D-72076 T\"ubingen, Germany
}

\author{K. Vogel}
\author{O. Crasser}
\author{R. Walser}
\author{W. P. Schleich}
\affiliation{
 Universit\"at Ulm,
 Abteilung Quantenphysik,
 D-89069 Ulm, Germany
}

\author{D.~Koelle}
\author{R.~Kleiner}
\affiliation{
  Physikalisches Institut II,
  Universit\"at T\"ubingen,
  Auf der Morgenstelle 14,
  D-72076 T\"ubingen, Germany
}



\begin{abstract}

  We consider a system of two semifluxons of opposite polarity in a 0-$\pi$-0 long Josephson junction, which classically can be in one of two degenerate states: \state{sa} or \state{as}. When the distance $a$ between the 0-$\pi$ boundaries (semifluxon's centers) is a bit larger than the crossover distance $a_c$, the system can switch from one state to the other due to thermal fluctuations or quantum tunneling. We map this problem to the dynamics of a single particle in a double well potential and estimate parameters for which quantum effects emerge. We also determine the classical-to-quantum crossover temperature as well as the tunneling rate (energy level splitting) between the states \state{sa} and \state{as}.
  
\end{abstract}

\maketitle

\section{Introduction}

$\pi$-Josephson junctions ($\pi$-JJs)\cite{Bulaevskii:pi-loop} are intensively investigated as they promise important advantages for Josephson junction based electronics\cite{Terzioglu:1997:CompJosLogic,Terzioglu:1998:CJJ-Logic}, and, in particular, for JJ based qubits\cite{Ioffe:1999:sds-waveQubit}. Nowadays a variety of technologies allow to manufacture such junctions\cite{Smilde:ZigzagPRL,Ryazanov:2001:SFS-PiJJ,Kontos:2002:SIFS-PiJJ,Tsuei:Review,Lombardi:2002:dWaveGB}.

One can also fabricate so-called long Josephson 0-$\pi$-junctions (0-$\pi$-LJJs)\cite{Bulaevskii:0-pi-LJJ}, \ie, LJJs some parts of which behave as 0-junctions and other parts as $\pi$-junctions. The most interesting fact about such junctions is that a vortex of supercurrent, carrying \emph{one half} of the magnetic flux quantum $\Phi_0\approx2.07\times10^{-15}\,{\rm Wb}$, can be formed at the boundaries between 0 and $\pi$ regions. Classically, this so-called semifluxon\cite{Goldobin:SF-Shape,Xu:SF-Shape}  has a degenerate ground state of either positive or negative polarity. The difference between positive and negative polarity is in the direction of the circulation of the supercurrent and, therefore, in the direction of the resulting magnetic field. The classical properties of semifluxons are under intense theoretical and experimental investigations\cite{Kogan:3CrystalVortices,Kirtley:SF:HTSGB,Kirtley:SF:T-dep,Hilgenkamp:zigzag:SF,Kirtley:IcH-PiLJJ,Goldobin:SF-ReArrange,Stefanakis:ZFS/2,Zenchuk:2003:AnalXover,Goldobin:Art-0-pi,Susanto:SF-gamma_c,Goldobin:2KappaGroundStates,Goldobin:F-SF}. While the classical properties of semifluxons (at least for systems with few semifluxons) are more or less understood, their quantum behavior and their possible applications in the quantum domain still have to be studied.

When the energy barrier separating two degenerate classical states is very small, the system may spontaneously switch from one state to the other due to thermal excitation over the barrier or due to quantum tunneling through the barrier. Thermally induced flipping of a single semifluxon was already observed\cite{Kirtley:SF:T-dep}. The quantum tunneling in the system of two coupled semifluxons was investigated theoretically by Kato and Imada\cite{Kato:1997:QuTunnel0pi0JJ} for the case of a biased junction, \ie, when the ground states are not degenerate and escape takes place in a certain direction. In view of possible applications and fundamental studies, it is interesting to see how degenerate semifluxon systems behave when the quantum effects start to exhibit themselves. 

In this paper, we study the two simplest systems: (a) one semifluxon with degenerate states \state{s} and \state{a} and (b) two coupled, antiferromagnetically (AFM) arranged semifluxons with degenerate states \state{sa} and \state{as}. In the first case, we use simple arguments to show that a single semifluxon is always deep in the classical limit. For a system of two semifluxons, we map the problem to the dynamics of a single particle in a double well potential and estimate relevant parameters for emergence of quantum effects. We also estimate the crossover temperature as well as the tunneling rate (energy level splitting) between the states \state{sa} and \state{as}.

\section{Model}

We consider a long one dimensional Josephson junction where the Josephson phase $\mu(x,t)$ is a continuous function of the coordinate $x$ along the LJJ and of time $t$. The dynamics of such a system is described by a Lagrangian $L=K-U$, where
\begin{equation}
  K = E_J \int_{-\infty}^{+\infty}  
    \omega_p^{-2}\frac{\mu_t^2}{2} 
  \,dx
  , \label{Eq:K}
\end{equation}
represents the kinetic energy and

\begin{equation}
  U = E_J \int_{-\infty}^{+\infty}  \left\{
    \lambda_J^2\frac{\mu_x^2}{2}   
    + \left[ 1-\cos(\mu+\theta(x)) \right] \right\}\,dx
  , \label{Eq:U}
\end{equation}
is a potential energy. The subscripts $x$ and $t$ denote the partial derivatives with respect to coordinate and time, accordingly. In the above equations the three physical parameters are the Josephson energy per unit of junction length $E_J$, the Josephson penetration depth $\lambda_J$ and the Josephson plasma frequency $\omega_p$. The function $\theta(x)$ describes the position of 0- and $\pi$-regions along the junction. It is zero along 0-regions and is equal to $\pi$ along $\pi$-regions.

It is straightforward to derive the equations of motion for the Josephson phase from the Lagrangian using the Euler-Lagrange prescription. Thus, on the classical level, one finds that the dynamics of the Josephson phase is described by the time-dependent sine-Gordon equation\cite{Goldobin:SF-Shape}
\begin{equation}
  \lambda_J^2\mu_{xx} - \omega_p^{-2} \mu_{tt} - \sin[\mu+\theta(x)] = 0
  . \label{Eq:sG:time}
\end{equation}
Here, damping and bias current are absent because dissipation and driving are not included in the initial Lagrangian. For the present discussion these terms are not required as we consider undriven dissipationless systems.

\section{A single semifluxon in a 0-$\pi$-junction}

Let us consider an infinite 0-$\pi$ LJJ. In this case $\theta(x)$ is a step function
\begin{equation}
  \theta(x) = 
    \left\{
      \begin{aligned}
       0,   &\quad x<0;\\
       \pi, &\quad x>0.
      \end{aligned}
    \right.
   \label{Eq:theta1}
\end{equation}

Classically, the ground state of this system is a single semifluxon.\cite{Bulaevskii:0-pi-LJJ,Xu:SF-Shape,Goldobin:SF-Shape,Goldobin:SF-ReArrange} Such a semifluxon may have positive or negative polarity that corresponds to two classical degenerated states \state{s} and \state{a}.\cite{Goldobin:SF-Shape,Goldobin:SF-ReArrange}

\begin{figure}[!tb]
  \includegraphics{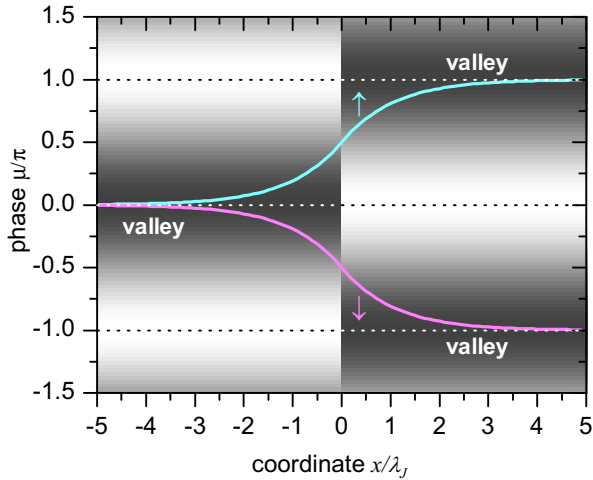}
  \caption{%
    Two solutions $\mu_{\protect\state{s}}(x)$ and $\mu_{\protect\state{a}}(x)$ corresponding to the two different states \protect\state{s} and \protect\state{a} of a single semifluxon in a 0-$\pi$-LJJ. The background color shows the corresponding Josephson potential energy density ${\cal U}=1-\cos(\mu+\theta)$, black corresponds to the valleys ${\cal U}=0$, while white corresponds to the summits ${\cal U}=2$.
  }
  \label{Fig:u-d}
\end{figure}

Each semifluxon's ground state can be considered as a string $\mu(x)$ laying in the potential profile ${\cal U}(x,\mu)=1-\cos[\mu+\theta(x)]$ (Josephson energy density). This potential profile looks like two sets of parallel valleys shifted relative to each other along $\mu$ direction by $\pi$ at $x=0$, see Fig.~\ref{Fig:u-d}. At $x\to-\infty$ the semifluxon's string is located in the valley $\mu=0$ and at $x\to+\infty$ it lays in the valley $\mu=\pm\pi$ for the states \state{s} and \state{a}, respectively. This is  shown in Fig.~\ref{Fig:u-d} by two curves corresponding to the states \state{s} and \state{a}. It is obvious that for quantum tunneling between \state{s} and \state{a} states the semi-infinite half of the string $x>0$ should tunnel from one valley to the other over the potential barrier with the height (per unit of junction length) of the order of $2E_J$. For finite LJJ, the energy barrier scales proportionally to the LJJ's length $L$, \ie, $\Delta U \sim E_J L$ and the probability of tunneling is exponentially small. For typical parameters (see discussion in Sec.~\ref{Sec:Est}) the thermal escape exponent $\Delta U/k_BT \gtrsim 100 L/\lambda_J$ at $T=4\units{K}$ and it becomes even larger at smaller $T$. The quantum escape exponent $\Delta U/\hbar\omega_0 \sim 400$, where $\omega_0\sim\omega_p$ is the eigenfrequency of a semifluxon.

The large barrier height results from the large length of the LJJ. As a modification, one can consider a LJJ of finite, rather small length $L<\lambda_J$. In this case the barrier height is finite and approaches zero when the junction length $L\to0$. In this limit, one can not really speak about a semifluxon. The solution for the phase in such a short junction can be found using the image technique (see Fig.~6 of Ref.~\onlinecite{Goldobin:Art-0-pi}). It represents a fragment of an infinite chain of antiferromagnetically (AFM) arranged semifluxons. The flux $\Phi$ present in the junction is much smaller than $\Phi_0/2$. The possibility of quantum tunneling in such a chain or in its fragments will be studied elsewhere.

For $L>\lambda_J$, another option for flipping from state \state{s} to \state{a} is the process of emitting a fluxon \state{F}, \ie, $\state{s}=\state{a}+\state{F}$. This process takes place already on the classical level\cite{Goldobin:SF-ReArrange,Goldobin:F-SF} and one does not need infinite energy to put the string from one valley to the other. However, it still requires a rather high energy $\sim 8E_J\lambda_J$ to create a fluxon. Consequently, this scenario will not be realized either.

If we consider the discontinuous Josephson phase $\phi(x,t)=\mu(x,t)+\theta(x)$\cite{Goldobin:SF-Shape} instead of the continuous phase $\mu$, the problem outlined above remains, but the semi-infinite tail should tunnel between $\phi(+\infty)=0$ and $\phi(+\infty)=2\pi$. 

Thus, we have shown that a single semifluxon in a LJJ is always in the classical limit if $L \gtrsim \lambda_J$. Therefore, below we consider the  more complex system of two coupled semifluxons, in which the barrier separating two classical states can be made quite small and quantum effects may emerge.

\section{Two coupled semifluxons in 0-$\pi$-0 junction}

\begin{figure}[!tb]
  \includegraphics{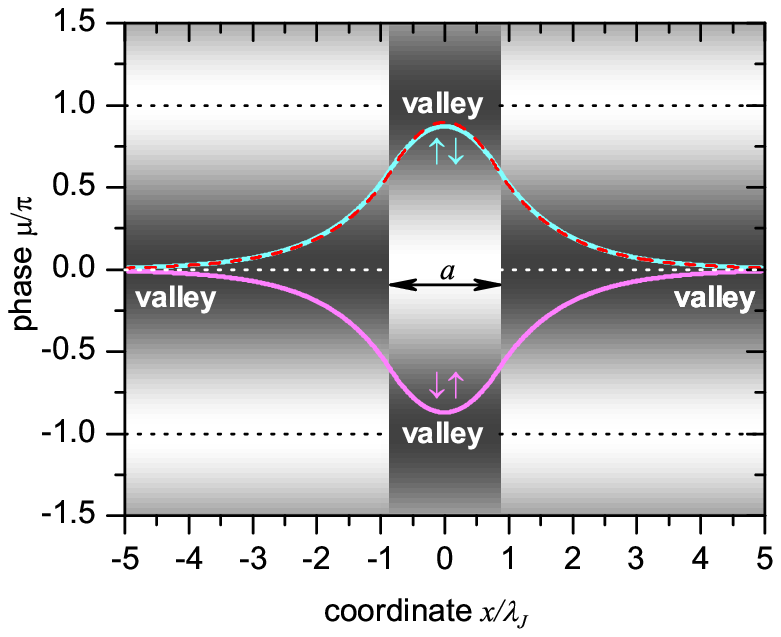}
  \caption{%
    Two solutions $\mu_{\protect\state{sa}}(x)$ and $\mu_{\protect\state{as}}(x)$ corresponding to the two different states \protect\state{sa} and \protect\state{as} of two semifluxons in 0-$\pi$-0-LJJ. The background color shows the corresponding Josephson potential energy ${\cal U}=1-\cos(\mu+\theta)$, black corresponds to the valleys ${\cal U}=0$, while white to the summits ${\cal U}=2$. The solid line shows the exact numerical solution of sine-Gordon Eq.~(\ref{Eq:sG:time}), while the dashed line (almost undistinguishable from the solid line) shows the approximate solution (\ref{Eq:mu0}) for $a=1.7\lambda_J$ for $B=+B_0$ (state \protect\state{sa}). 
  }
  \label{Fig:ud-du}
\end{figure}

Let us consider an infinite 0-$\pi$-0 LJJ. In this case 
the function $\theta(x)$ is a step function with $+\pi$ and $-\pi$ discontinuities situated at $x=\pm a/2$:
\begin{equation}
  \theta(x) = 
    \left\{
      \begin{aligned}
       0,   &\quad |x|>\frac{a}{2};\\
       \pi, &\quad |x|<\frac{a}{2},
      \end{aligned}
    \right.
  \label{Eq:theta}
\end{equation}
where $a$ is the length of the $\pi$-region between 0-$\pi$ boundaries. The ground state in such a junction crucially depends on $a$. If the distance $a$ is smaller than a crossover distance $a_c=(\pi/2)\lambda_J$, the ground state of the system is the so-called flat phase state $\mu(x) \equiv 0$, while for $a>a_c$ two antiferromagnetically (AFM) ordered semifluxons form the ground state.\cite{Kato:1997:QuTunnel0pi0JJ,Goldobin:SF-ReArrange,Zenchuk:2003:AnalXover} Due to symmetry reasons, there are two possible semifluxon states \state{sa} and \state{as} that have the same energy, see Fig.~\ref{Fig:ud-du}. We would like to calculate the tunneling probability or the energy level splitting due to the coupling between these two states. 

For $a>a_c$ appreciable tunneling takes place, if the energy barrier between the states \state{sa} and \state{as} is small. It is the case when the distance $a$ is a bit larger than the crossover distance $a_c$, \ie, $a=a_c+\delta a$, $\delta a\ll\lambda_J$. The energy of the unstable flat phase state plays a role of a potential barrier. When the distance $a \to a_c$, the barrier disappears and both states turn into a flat phase state\cite{Kato:1997:QuTunnel0pi0JJ,Goldobin:SF-ReArrange,Zenchuk:2003:AnalXover}.

\subsection{Collective coordinate}

In this subsection, we analyze the static solutions of the sine-Gordon equation corresponding to the \state{sa} or \state{as} state and derive a simple analytic approximation for them in the limit $\delta a\ll\lambda_J$. This analytic solution has an amplitude parameter $B$, which can be used as a collective coordinate to map our problem onto the dynamics of a fictitious single particle in an effective potential.

To analyze the static solutions we assume that $\mu$ does not depend on  $t$, \ie, the second term in Eq.~(\ref{Eq:sG:time}) vanishes. In the limit $\da=\delta a/\lambda_J \ll 1$, \ie, when the states \state{sa} and \state{as} are very similar to the flat phase state $\mu=0$, we solve the stationary sine-Gordon equation in the regions 1 ($x < -a/2$), 2 ($-a/2 < x < a/2$) and 3 ($x > a/2$), assuming that the phase $\mu(x) \ll 1$. In the regions 1 and 3 we find $\mu_{1,3}(x) = A_{1,3} e^{-|x|/\lambda_J}$ ($A_{1,3}\ll1$), while in the region 2 we have $\mu_{2}(x) = B \cos(x/\lambda_J)$ ($B\ll1$). By matching the boundary conditions $\mu_1(-a/2) = \mu_{2}(-a/2)$ and $\mu_{2}(a/2) = \mu_{3}(a/2)$, we can express $A_{1,3}$ in terms of $B$, namely $A_{1,3}=B\cos({a}/{2\lambda_J}) \exp(a/2\lambda_J)$. Thus we finally arrive at
\begin{equation}
  \mu(x) =
  B \left\{
    \begin{aligned}
     &\cos\left( \frac{a}{2\lambda_J} \right) 
     e^{{a}/{2\lambda_J}} e^{{-|x|}/{\lambda_J}}
     , & |x|>\frac{a}{2} ;\\
     &\cos\left( \frac{x}{\lambda_J} \right)
     , & |x|<\frac{a}{2} .     
    \end{aligned}
  \right.
  \label{Eq:mu0}
\end{equation}
Note, that solution (\ref{Eq:mu0}) does not satisfy the condition for continuity of the derivative $\mu_x(\pm a/2)$. The mismatch in derivatives is rather small $(\sim B\da)$ and is beyond the order of our approximation. 

In Fig.~\ref{Fig:ud-du} we compare the exact numerical solution with $\mu(x)$ given by formula (\ref{Eq:mu0}) for $a=1.7\lambda_J$ ($\da\approx0.13$). We see that they are almost indistinguishable. In principle, Eq.~(\ref{Eq:mu0}) satisfies the linearized sine-Gordon equation for any $B\ll1$. However, as we show later, only two values $B=\pm B_0$ minimize the energy of the system. By changing the parameter $B$, we can make a smooth transition between the ground states \state{sa} and \state{as}. Thus, we can use $B$ as a collective coordinate. One can think about a fictitious particle with an effective mass $M$, which moves along the coordinate $B$ in an effective potential $U(B)$ derived below in Sec.~\ref{Sec:TwoSemifluxons:U}.

To describe the \emph{dynamics} of this particle, we let the $B$ become a dynamic variable $B(t)$. Thus, the shape of the solution $\mu(x)$ is fixed, but its amplitude depends on time. In this way $B(t)$ approximately describes the dynamics of our system and, in particular, transition between states. 

One can view the derivation of the approximate solution and the following introduction of the collective coordinate from different angle.  For $a < a_c$, the flat phase state $\mu(x) \equiv 0$ is a stable solution of the static sine-Gordon equation and all linear response eigenmodes of the  time-dependent sine-Gordon equation (\ref{Eq:sG:time}) are stable. For $a > a_c$ the lowest eigenmode becomes unstable. Therefore, within a linear approximation this mode would grow exponentially. Since the sine-Gordon equation is a nonlinear equation, the mode amplitude $B$ would not become arbitrarily large, but saturates at some value $B=B_0$. Basically, this represents a new solution (\ref{Eq:mu0}) for $a>a_c$. Then we assume that the amplitude $B$ of this mode is large in comparison with amplitudes of all other modes, but still small enough so that we only have to take into account terms up to the order $B^4$ when we will calculate the energy in Sec.~\ref{Sec:TwoSemifluxons:U}. All other modes are neglected. A similar approach was used by Kato and Imada\cite{Kato:1997:QuTunnel0pi0JJ}. This assumption essentially means that Eq.~(\ref{Eq:mu0}) is a good approximation for $\mu(x)$ for $\da \ll 1$. Consequently, we use the lowest mode amplitude $B$ as a collective coordinate. Such an approximation is justified because this lowest mode, according to the Sturm-Liouville theorem, has no zeros.

Thus, instead of treating $\mu(x,t)$ as a classical (quantum) field we can restrict ourselves to single particle classical (quantum) dynamics in a one dimensional potential.

\subsection{Effective mass and potential}
\label{Sec:TwoSemifluxons:U}

To determine the effective mass $M$ associated with the collective coordinate $B(t)$ we substitute our solution (\ref{Eq:mu0}) into the kinetic energy Eq.~(\ref{Eq:K}). After integration we obtain
\begin{equation}
  \begin{split}
    K(\dot{B}) 
    &= \frac{E_J\lambda_J}{4\omega_p^2}
      \left[ 1+\tilde{a}+\cos(\tilde{a})+\sin(\tilde{a}) \right]\dot{B}^2(t)\\
    &\approx \frac{E_J\lambda_J}{8\omega_p^2}
      \left( 4+\pi\right)\dot{B}^2(t)
    , \label{Eq:K(B)}
  \end{split}
\end{equation}
where $\tilde{a}=a/\lambda_J$. When we compare this result to the standard expression $K(\dot{B})=M\dot{B}^2/2$ for the kinetic energy, we find that the inertial mass of a particle is given by
\begin{equation}
  M \approx \frac{E_J\lambda_J}{4\omega_p^2} (4+\pi)
  . \label{Eq:M} 
\end{equation}

Similarly, we find the effective potential $U(B)$ by substituting the solution (\ref{Eq:mu0}) into Eq.~(\ref{Eq:U}). To calculate the integral in terms of elementary functions, we assume that $B$ is small and expand the integrand up to terms $\sim B^4$.

\begin{figure}[tb]
  \centering\includegraphics{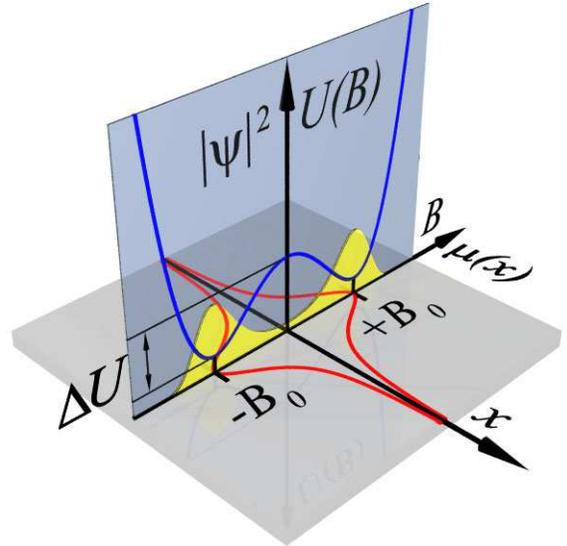}
  \caption{%
    Schematic view on two classically stable states \protect\state{sa} and \protect\state{as} and the corresponding mapping to a single particle moving along coordinate $B$ in a one dimensional potential $U(B)$. Quantum mechanically one should speak about probability density $|\psi(B)|^2$ to find the particle at different locations (area plot). 
  }
  \label{Fig:x-phi-U}
\end{figure}

For small $\da$, we obtain the following expression 

\begin{equation}
  U(B,\da)\approx E_J\lambda_J\left[ 
    \frac{\pi+2}{128}B^4 - \frac12 \da B^2 
    + 2\da + \pi
  \right].
  \label{Eq:U(B)}
\end{equation}
One can see that for a given distance $\da$, the potential energy $U(B)$ is a double well potential with walls $\propto B^4$ and an energy barrier in the middle $\propto -B^2$, see Fig.~\ref{Fig:x-phi-U}. The potential energy has two minima at $B=\pm B_0$, where
\begin{equation}
  B_0 = \sqrt{\frac{32}{\pi+2}}\sqrt{\da}
  . \label{Eq:B0}
\end{equation}
%
A similar result was obtained earlier\cite{Kato:1997:QuTunnel0pi0JJ,Kato:Bug}. Thus, the two classically stable solutions \state{sa} and \state{as} correspond to $B=\pm B_0$ and can be seen on $x$--$\mu$ plane in Fig.~\ref{Fig:x-phi-U}. The corresponding potential $U(B)$ is shown on $U$--$B$ plane.

After we have determined the mass $M$ and the effective potential $U(B)$ we can calculate two parameters which are important for the behavior of the system: the height $\Delta U$ of the energy barrier in $U(B)$ and the frequency $\omega_0$ associated with small harmonic oscillations around $B=\pm B_0$. The height of the energy barrier is given by
\begin{equation}
   \Delta U(\da) = U(0,\da) - U(B_0,\da) =
   \frac{8}{\pi+2} E_J \lambda_J \da^2.
   \label{Eq:DeltaU}
\end{equation}

The harmonic oscillator frequency will be the same for small oscillations around $B_0$ and $-B_0$ since the potential $U(B)$ is symmetric. It is therefore sufficient to consider small oscillations around $B_0$. In order to find the harmonic oscillator frequency $\omega_0$ we expand the potential $U(B)$ around $B_0$ and find
\begin{equation}
  U(B_0+\delta B) \approx E_J\lambda_J \left[
    \left( \pi+2{\da}-\frac{8{\da}^2}{\pi+2} \right)
    + \underbrace{\da {\delta B}^2}_{\text{one well}}
  \right]
  , \label{Eq:U@min}
\end{equation}
where we have neglected terms proportional to $\da^3$ and $\da^4$.
Therefore, $\omega_0$ is given by
\begin{equation}
\omega_0= \sqrt{\frac{2 E_J \lambda_J}{M}}\sqrt{\da}
  = \sqrt{\frac{8}{\pi+4}} \omega_p \sqrt{\da} .
  \label{Eq:omega_0}
\end{equation}

To check our analytical expressions derived above, we have compared the stationary numerical solution of the full sine-Gordon Eq.~(\ref{Eq:sG:time}) to the solution (\ref{Eq:mu0}) with $B=B_0$ from Eq.~(\ref{Eq:B0}), see Fig.~\ref{Fig:ud-du}. We have also used the stationary solution of the full sine-Gordon equation to calculate the energy difference $\Delta U(\da)$ between the flat phase state and the state $\state{sa}$ and compared it to the energy barrier $\Delta U(\da)$ given in Eq.~(\ref{Eq:DeltaU}). Furthermore, we have calculated the lowest eigenfrequency $\omega_0(\da)$ of the state $\state{sa}$ and compared it with Eq.~(\ref{Eq:omega_0}). These simulations were done for $a$ in the range $1.57\lambda_J \ldots 2.00\lambda_J$ ($\da=0\ldots0.43$). For all three quantities we found excellent agreement between analytical expressions and numerical results in the limit $\da\to0$. Even for $a=2\lambda_J$ ($\da=0.43$) the discrepancy between analytics and numerics is $\sim 9\,{\rm \%}$ for $\mu(0)$, $\sim31\,{\rm \%}$ for $\Delta U$ and $\sim 21\,{\rm \%}$ for $\omega_0$. Thus, our analytical approximation (\ref{Eq:mu0}) with the collective coordinate $B$ describes the classical dynamics of our system in the limit $\da\ll1$ quite well. This suggests that we can also successfully use this collective coordinate approach for a quantum mechanical description.

Before we introduce the Schr\"odinger equation for the system, we briefly summarize the classical dynamics of the system: The two stationary solutions $\pm B_0$ correspond to the \state{sa}-state and the \state{as}-state, see Fig.~\ref{Fig:x-phi-U}. The position dependence is taken into account by Eq.~(\ref{Eq:mu0}). These two solutions with $B=\pm B_0$ minimize the energy of the system. For energies smaller than the barrier height $\Delta U$, $B$ is restricted to one of the potential wells of $U(B)$, Eq.~(\ref{Eq:U(B)}), and will move between two turning points. For sufficiently small energies $B(t)$ describes harmonic oscillations with a frequency $\omega_0$ given by Eq.~(\ref{Eq:omega_0}).

\subsection{Schr\"odinger equation}

According to the classical picture, our system has two stable states \state{sa} and \state{as} corresponding to $B = \pm B_0$, \ie, a single particle in one of the wells of the potential $U(B)$ (\ref{Eq:U(B)}). Now, we consider the problem quantum mechanically and ask what is the probability that the particle tunnels, \eg, between the state corresponding to the classical positions $-B_0$ and $B_0$. In quantum mechanics our collective coordinate becomes an operator. In ``position'' representation (or $B$-representation) the stationary Schr\"odinger equation reads
\begin{equation}
  \left[ -\frac{\hbar^2}{2M}\fracp[2]{}{B} + U(B) \right] \psi(B) = E \psi(B)
  , \label{Eq:Schroedinger}
\end{equation}
where $M$ is the effective mass defined by Eq.~(\ref{Eq:M}) and $U(B)$ is the effective potential given by Eq.~(\ref{Eq:U(B)}).

To gain more insight, we measure the energy $E$ in units of $E_J \lambda_J$. Using the expression for the mass $M$, Eq.~(\ref{Eq:M}), we can write the Schr\"odinger equation (\ref{Eq:Schroedinger}) in the form
\begin{equation}
  \left[ -\frac{2}{\pi+4}
     \left(\frac{\hbar\omega_p}{E_J \lambda_J}\right)^2
        \fracp[2]{}{B} + u(B) \right] \psi(B)
  = \varepsilon \psi(B),
  \label{Eq:Schroedinger-scaled}
\end{equation}
where the energy eigenvalue $\varepsilon$ is defined by $\varepsilon = E/(E_J \lambda_J)$ and
\begin{equation}
  u(B) = \frac{U(B)}{E_J \lambda_J}
  \approx \left[ 
    \frac{\pi+2}{128}B^4 - \frac12 \da B^2 
    + 2\da + \pi
  \right]
  \label{Eq:u}
\end{equation}
is the scaled effective potential. There are only two dimensionless parameters in the scaled Schr\"odinger equation (\ref{Eq:Schroedinger-scaled}): the distance between semifluxons $\da$ and the dimensionless factor $\hbar \omega_p/(E_J \lambda_J)$. This factor plays the role of a scaled $\hbar$ and quantum effects will vanish in the limit $\hbar \omega_p/(E_J \lambda_J) \to 0$.

In section \ref{Sec:QT}, we will investigate quantum tunneling based on the Schr\"odinger equation (\ref{Eq:Schroedinger}). Before we start with these calculations we would like to make some simple estimations to see whether we can expect to observe quantum effects in our system.

\subsection{Estimation of quantum-to-classical crossover}
\label{Sec:Est}

For a harmonic oscillator with mass $M$ and frequency $\omega_0$ the width of the ground state is determined by
\begin{equation}
  \langle x^2 \rangle = \frac{\hbar}{M\omega_0}.
  \label{Eq:HO-width}
\end{equation}
Our potential $U(B)$, Eq.~(\ref{Eq:U(B)}), is not harmonic, but for sufficiently large energy barriers we may approximate each potential well by an harmonic oscillator, see Eq.~(\ref{Eq:U@min}). We can then use Eq.~(\ref{Eq:HO-width}) to estimate the spread of a wave function in each of the potential wells. Using our expressions for the mass $M$ (\ref{Eq:M}) and the frequency $\omega_0$ (\ref{Eq:omega_0}), we find
\begin{equation}
  \langle {\delta B}^2 \rangle = 
  \frac{2}{\sqrt{4+\pi}}\frac{\hbar\omega_p}{E_J\lambda_J}
  \frac{1}{\sqrt{\da}}
  . \label{Eq:B2av}
\end{equation}
Quantum effects are noticeable when the wave function in the left well overlaps with the wave function of the right well. This overlap should be appreciable, but not too large since otherwise two states will not be distinguishable anymore. For rough estimation we take   $\langle {\delta B}^2 \rangle \gtrsim 0.1 B_0^2$ as criterion for quantum behavior. Thus quantum effects will dominate if [we use $B_0$ from Eq.~(\ref{Eq:B0})]
\begin{equation}
  \begin{split}
    \frac{\langle {\delta B}^2 \rangle}{B_0^2}
    &=\frac{\pi+2}{16\sqrt{4+\pi}}
    \frac{\hbar\omega_p}{E_J\lambda_J} \frac{1}{{\da}^{3/2}} \\
    &= \frac{\pi+2}{16\sqrt{4+\pi}} 
    (2e)^2\sqrt{\frac{2\mu_0\lambda_L}{C w^2}} 
    \frac{1}{{\da}^{3/2}}\gtrsim 0.1
    , \label{Eq:QuDomination}
  \end{split}
\end{equation}
where we took into account the definitions
\begin{equation}
  \lambda_J = \sqrt{\frac{\Phi_0}{2\pi\mu_0 d' j_c}},\;
  \omega_p  = \sqrt{\frac{2\pi j_c}{\Phi_0 C}},\;
  E_J       = \frac{j_c w \Phi_0}{2\pi}
  . \label{Eq:def}
\end{equation}
In Eq.~(\ref{Eq:def}), $\mu_0d'$ is the inductance per square of the superconducting electrodes ($\mu_0$ is a permeability of vacuum, $d' \approx 2\lambda_L$, $\lambda_L$ is the London penetration depth), $j_c$ is the critical current density of the LJJ, $C$ is the capacitance of the LJJ per unit of area and $w$ is the LJJ's width.

For typical parameters $\lambda_L=100\units{nm}$, $w=1\units{\mu m}$ and $C=4.1\units{\mu F/cm^2}$ (\textsc{Hypres}\cite{Hypres} technology with $j_c=100 \units{A/cm^2}$) we get 
\begin{equation}
  \frac{\langle {\delta B}^2 \rangle}{B_0^2}
  \approx 3\times10^{-4}{\da}^{-3/2}\gtrsim 0.1
  . \label{Eq:EstSpread}
\end{equation}

Thus, quantum effects start to play a role for ${\da}\lesssim0.02$. It is interesting to note that, according to Eq.~(\ref{Eq:QuDomination}), the occurrence of quantum effects \emph{does not depend on $\hbar$!}\cite{HiddenHbar}

Using definitions (\ref{Eq:def}) in terms of physical parameters of LJJ, we can express inertial mass $M$ (\ref{Eq:M}) as
\begin{equation}
  M \approx \frac{(4+\pi)wC}{4\sqrt{\mu_0 d' j_c}}
  \left( \frac{\Phi_0}{2\pi} \right)^\frac52
  \approx 2.4\times10^{-4}m_e \lambda_J^2
  . \label{Eq:M:ph}
\end{equation}

To estimate the crossover temperature, we compare the barrier height $\Delta U$ (\ref{Eq:DeltaU}) with $k_B T$, \ie,
\begin{equation}
  T^{\star} = \frac{\Delta U}{k_B}
  = \frac{E_J\lambda_J}{k_B}\frac{8}{\pi+2}\da^2
  . \label{Eq:T*}
\end{equation}
For $\da=0.01$, we obtain $T^\star\approx 130\units{mK}$, which is a reasonable value for observation using modern $^3$He/$^4$He dilution refrigerators. This value is also typical for other types of qubits based on JJs.\cite{Ustinov:FluxonQuantumTunneling,Mooij:1999:JosPersistCurrentQubit,Wal:2000:QuSuperposPersCurrStat,Grajcar:2004:FluxQubitTunnelAmp} We would like to point out that $T^\star \propto w$ via $E_J$. This is natural since it is rather difficult to thermally activate a ``heavy'' vortex. 

\section{Quantum tunneling}
\label{Sec:QT}

Quantum mechanics tells us that if we start with a wave function which is localized at one of the two minima of the potential $U(B)$ it will tunnel through the barrier and therefore also populate the other minimum. We could follow this picture by using a wave function which is localized in one of the minima as an initial condition and solve the time-dependent version of the Schr\"odinger equation (\ref{Eq:Schroedinger}) numerically. Obviously, in this approach the answer is not a single number since details will depend on the exact form of the initial wave function.

\begin{figure}
  \includegraphics[width=0.9\columnwidth]{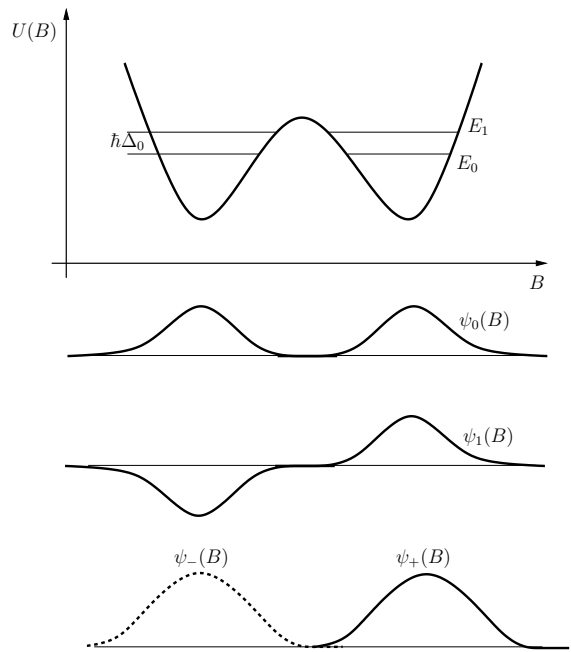}
  \caption{
    The two lowest energy eigenvalues in a double well potential and the corresponding eigenfunctions. The ground state $\psi_0(B)$ is symmetric whereas the first excited state $\psi_1(B)$ is anti-symmetric. The sum $\psi_{+}(B)$ and the difference $\psi_{-}(B)$ of these two energy eigenfunctions are localized in the right or in the left well, respectively.
  }
  \label{Fig:psi-sketch}
\end{figure}

We use a different approach which is based on the following picture: Let us assume for the moment that the two wells of $U(B)$ are separated by a sufficiently high energy barrier. Then two lowest energy eigenvalues $E_0$ and $E_1$ differ by the tunnel splitting $\hbar\Delta_0=E_1-E_0$ which is small compared to $\hbar\omega_0$. The corresponding eigenfunctions are  $\psi_0(B)$ and $\psi_1(B)$. The ground state $\psi_0(B)$ is symmetric whereas the first excited state $\psi_1(B)$ is anti-symmetric. We now use the sum and the difference to define $\psi_{\pm}$
\[
  \psi_{\pm}(B) = \frac{1}{\sqrt{2}} 
  \left[ \psi_0(B) \pm \psi_1(B) \right]
\]
of the two eigenfunctions as an initial condition. These two wave functions are well localized in one of the wells of $U(B)$, see Fig.~\ref{Fig:psi-sketch}. The time evolution of $\psi_{\pm}(B)$ is given by
\begin{equation}
  \begin{split}
    \psi_{\pm}(B,t) 
    &= \frac{1}{\sqrt{2}} \left[
       \psi_0(B) e^{-\frac{i}{\hbar} E_0 t}
       \pm \psi_1(B) e^{-\frac{i}{\hbar} E_1 t}
    \right]\\
    &= \frac{1}{\sqrt{2}} e^{-\frac{i}{\hbar} E_0 t}\left[
       \psi_0(B) \pm \psi_1(B) e^{-i\Delta_0 t}
    \right].
  \end{split}
  \label{Eq:psi-pm}
\end{equation}
As we can see, $\psi_{\pm}(B,t)$ coherently oscillates between $\psi_{+}(B)$ and $\psi_{-}(B)$ (apart from an overall phase factor). The oscillation frequency is given by $\Delta_0$.

Our calculations are still correct if the energy barrier becomes smaller (or even disappears). However, the wave functions $\psi_{+}(B)$ and $\psi_{-}(B)$ are not well localized in one of the wells of $U(B)$ anymore. Instead they will significantly extend into the other well. Nevertheless, the oscillation frequency between the states $\psi_{+}(B)$ and $\psi_{-}(B)$ is still given by $\Delta_0$, although the interpretation that the oscillations between these two states describe a tunneling process through a barrier may become questionable at some point. In the rest of this section we will calculate the energy splitting $\hbar\Delta_0$. 


\subsection{Semiclassical limit}

In the semiclassical limit the tunnel splitting $\hbar\Delta_0$ can be calculated using standard methods like WKB \cite{LandauLifshitz3} or the instanton technique \cite{Kleinert:PathInt,Weiss:DissQuSys} to obtain analytical results. In the latter the problem of finding $\hbar\Delta_0$ is essentially reduced to determine the classical path which connects the two maxima of the inverted potential $-U(B)$ [the minima of $U(B)$] and calculate the corresponding action.
For a quartic double-well potential of the form (\ref{Eq:U(B)}) these calculations can be performed analytically. We find\cite{Weiss:DissQuSys}
\begin{equation}
  \Delta_0 = 8 \omega_0\, \sqrt{\frac{2\Delta U}{\pi \hbar\omega_0}}
  \, \exp \left\{-\frac{16\Delta U}{3\hbar\omega_0}\right\},
  \label{Eq:Delta-semiclassical-1}
\end{equation}
As we might have expected, $\Delta_0$ decreases exponentially with the barrier height $\Delta U$.

Using our expressions for $\Delta U$ (\ref{Eq:DeltaU}) and $\omega_0$ (\ref{Eq:omega_0}), we arrive at
\begin{equation}
  \delta\varepsilon=\frac{\hbar\Delta_0}{E_J \lambda_J} = 
  K_1\da^{\frac54}\sqrt{\frac{\hbar\omega_p}{E_J \lambda_J}}
  \exp \left(
     - K_2\frac{E_J \lambda_J}{\hbar\omega_p} \da^{\frac32}
   \right).
  \label{Eq:Delta-semiclassical-2}
\end{equation}
where the two numerical factors $K_1$ and $K_2$ are given by
\begin{eqnarray}
  K_1=\frac{32}{\sqrt{\pi(\pi+2)}}\left(\frac{8}{\pi+4}\right)^{1/4}
  \approx 8.2
  , \label{Eq:K1}\\
  K_2=\frac{32\sqrt{2(\pi+4)}}{3(\pi+2)}
  \approx 7.84 
  . \label{Eq:K2}
\end{eqnarray}
Since the factor $E_J \lambda_J/(\hbar\omega_p) \da^{3/2}$ appears in the exponent, the energy splitting $\hbar\Delta_0$ is very sensitive to this number. Please note, that the inverse of this factor appears in Eq.~(\ref{Eq:QuDomination}).

\subsection{Numerical results}

\begin{figure}
  \includegraphics{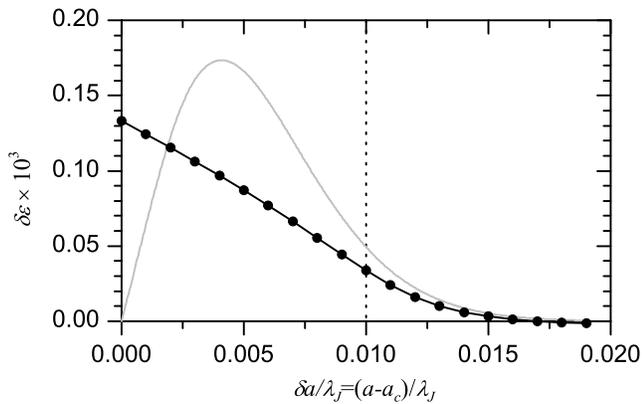}
  \caption{%
    The energy splitting $\delta\varepsilon$ as a function of $\da$ for $\hbar \omega_p/(E_J \lambda_J) = 2.4 \times 10^{-3}$. The gray line is the energy splitting according to the semiclassical approximation Eq.~(\ref{Eq:Delta-semiclassical-2}), while symbols show $\delta\varepsilon$ obtained by numerical solution of Eq.~(\ref{Eq:Schroedinger-scaled}).
  }
  \label{Fig:splitting}
\end{figure}

\begin{figure*}[!tb]
  \includegraphics{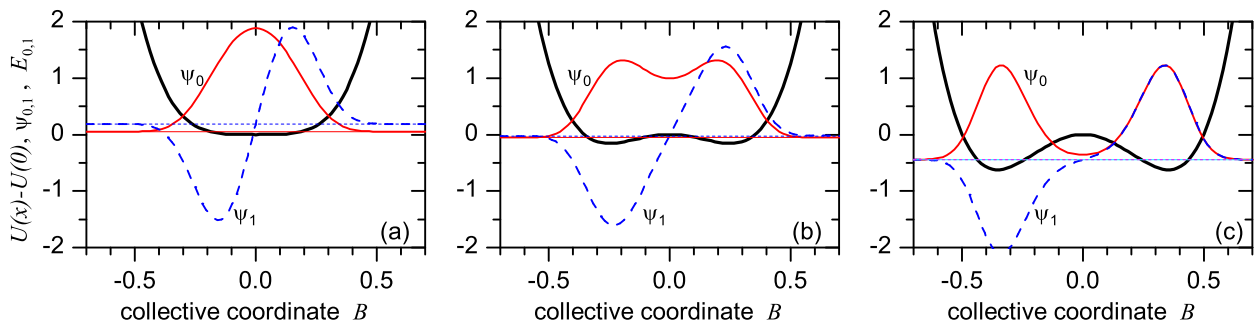}
  \caption{%
    Potential $U(\da)-U(0)$ (thick line), the two lowest eigenfunctions $\psi_{0,1}$ (solid and dashed lines) and the two corresponding lowest energy eigenvalues $E_{0,1}$ (thin horizontal solid and dashed lines) for (a) $\da=0$, (b) $\da=0.01$ and (c) $\da=0.02$. 
  }
  \label{Fig:eigenfns}
\end{figure*}

Our numerical calculations are based on the scaled Schr\"odinger equation (\ref{Eq:Schroedinger-scaled}). In Fig.~\ref{Fig:splitting} we have plotted the energy splitting $\delta\varepsilon$ as a function of $\da$ for a fixed value of $\hbar \omega_p/(E_J \lambda_J)=2.4 \times 10^{-3}$ which corresponds to the parameters used in our estimations in Sec.~\ref{Sec:Est}. According to Fig.~\ref{Fig:splitting}, our semiclassical expression (\ref{Eq:Delta-semiclassical-2}) describes the energy splitting reasonably well for $\da > 0.01 $. For the parameters used in Sec.~\ref{Sec:Est} our numerical calculations predict $\Delta_0/2\pi\approx 0.61\units{GHz}$ whereas the semiclassical formula (\ref{Eq:Delta-semiclassical-2}) gives $\Delta_0/2\pi \approx 0.88\text{~GHz}$.

The two lowest eigenfunctions of the Schr\"odinger Eq.~(\ref{Eq:Schroedinger-scaled}) are shown in Fig.~\ref{Fig:eigenfns} for three values of $\da$. For $\da=0$ two lowest states represent the ground and the first excited states of the particle in a $B^4$ potential with a relatively large spacing between the energy levels, see Fig.~\ref{Fig:eigenfns}(a). For $\da=0.02$, in Fig.~\ref{Fig:eigenfns}(c) one can see clearly that the wave functions become strongly localized in the potential minima. Therefore, the quantum tunneling is suppressed, see also Fig.~\ref{Fig:splitting}. Finally, for the case $\da=0.01$ shown in Fig.~\ref{Fig:eigenfns}(b), we have rather strong coupling and appreciable energy level splitting due to the wave functions overlap. The energy level splitting $\delta\varepsilon$ as function of $\da$ is plotted in Fig.~\ref{Fig:splitting}.

\section{Conclusion}

In conclusion, we have tried to map the problem of quantum evolution of the Josephson phase (quantum field theory) to the motion of a point-like particle in a double well potential. For the case of two coupled semifluxons arranged antiferromagnetically, we have estimated that the quantum effects start dominating when the length $a$ of the $\pi$-region exceeds $a_c=\frac{\pi}{2}\lambda_J$ by less then $0.02\lambda_J$, see Eq.~(\ref{Eq:EstSpread}). Corresponding frequency $\Delta_0$ of the wave function oscillation between the two states is $\sim 1\,{\rm GHz}$, which is good value to detect experimentally. The estimated classical-to-quantum crossover temperature $T^*\sim130\,{\rm mK}$ (\ref{Eq:T*}) lays in a range accessible for modern $^3$He/$^4$He dilution refrigerators and also represents the typical crossover temperature for flux and fluxon qubits\cite{Ustinov:FluxonQuantumTunneling,Mooij:1999:JosPersistCurrentQubit,Wal:2000:QuSuperposPersCurrStat,Grajcar:2004:FluxQubitTunnelAmp}.

We would like to mention that technology is advanced enough to fabricate huge arrays of 0-$\pi$-junctions carrying thousands of semifluxons\cite{Hilgenkamp:zigzag:SF}. Thus, in the future it will be interesting to extend the results obtained here to larger systems, \eg, to one or two dimensional fractional vortex crystals. 

On the other hand, it is also interesting to consider a single semifluxon squeezed into a rather short junction. In this case, the system is very similar to a flux qubit with zero loop area and flux conservation does not prevent flipping between the states \state{s} and \state{a}.

Not all problems can be so easily mapped to the single particle dynamics. For example, the experimentally relevant problem of quantum escape $\state{s}\to\state{a}+\state{F}$ at the semifluxon's depinning current $I\to\frac{\pi}{2}I_c$ probably will need more elaborate approaches. 

\acknowledgments

We express our sincere gratitude to T. Pfau who initiated this interdisciplinary collaboration as a coordinator of the SFB/TR21 proposal to Deutsche Forschungsgemeinschaft (DFG). W.P.S. and R.W. also gratefully acknowledge financial support from the Landesstiftung Baden-Wurttemberg (grant \#33854). E.G. acknowledges support of DFG (project GO-1106/1) and ESF program "PiShift".

\bibliography{this,jj-annular,LJJ,SFS,pi,SF,software,QuComp}

\begin{thebibliography}{35}
\bibitem[{Hid()}]{HiddenHbar}
\bibinfo{note}{In fact, an $\hbar$ is still hidden in $\lambda_J$
  (\ref{Eq:def}) of $\da=\delta a/\lambda_J$.}
\end{thebibliography}

\end{document}